\documentclass[usenatbib]{mn2e}
\usepackage{epsfig}

\def\beq{\begin{eqnarray}}
\def\eeq{\end{eqnarray}}
\def\delv{\Delta v}
\def\kms{\, \rm{km}\,  \rm{s}^{-1}}
\def\cm2{{\, \rm{cm}}^{-2}}
\def\vvir{v_{vir}}
\def\dNdz{\rm {{dN}/{dz}}}
\def\frto{f_{ratio}}
\def\fciv{f_{CIV}}

\title{Damped Lyman alpha systems and galaxy formation models - II.
High ions and Lyman limit systems}

\author[A. H. Maller, J. X. Prochaska, R. S. Somerville  and J. R. Primack]
{Ariyeh H. Maller$^1$, Jason X. Prochaska$^2$, Rachel S. Somerville$^3$ 
and Joel R. Primack$^4$\\
$^1$Astronomy Department, UMASS, Amherst, 01003\\
$^2$Observatories of the Carnegie Institution of Washington, 
Pasadena CA 91101\\
$^3$Astronomy Department, Univeristy of Michigan, Ann Arbor, 
MI 48109\\ 
$^4$Physics Department, University of California, Santa Cruz, 
CA 95064}
 
\begin{document} 

\maketitle
 
\begin{abstract} 
We investigate a model for the high-ionization state gas associated
with observed damped Lyman-$\alpha$ systems, based on a semi-analytic
model of galaxy formation set within the paradigm of hierarchical
structure formation. In our model, the hot gas in halos and sub-halos
is assumed to be in a multi-phase medium which gives rise to CIV
absorption, while the low-ionization state gas is associated with the
cold gas in galaxies.  The model matches the distribution of CIV
column densities if we assume that the hot gas has a mean metallicity
log C/H =$-1.5$, which is the observed mean metallicity of damped
systems. The same model then leads naturally to kinematic properties
that are in good agreement with the data, for both the low- and
high-ionization state gas.

We examine the contribution of both hot and cold gas to sub-damped
systems ($N_{HI} > 4 \times 10^{19} \cm2$) and suggest that the
properties of these systems can be used as an important test of the
model. We expect that sub-DLA systems will generally be composed of a
single gas disk and thus predict that they should have markedly
different kinematics than the damped systems.  We also find that the
frequency of absorbers drops dramatically for column densities below
$4 \times 10^{19} \cm2$.  These results are a consequence of our model
for damped Lyman-$\alpha$ systems and we believe they are a generic
prediction of multi-component models.

Finally, we find that hot halo gas produces less than one third of
Lyman limit systems at redshift three. We model the contribution of
mini-halos (halos with virial velocities $\le 35 \kms$) to Lyman limit
systems and find that they may contain as much gas as is observed in
these systems.  However, if we adopt realistic models of the gas
density distribution we find that these systems are not a significant
source of Lyman limit absorption.  Instead we suggest that uncollapsed
gas outside of virialized halos is responsible for most of the Lyman
limit systems at high redshift.
\end{abstract}


\section{Introduction}
\label{sec_int}

Semi-analytic models of galaxy formation, set within the hierarchical
paradigm of structure formation generic to Cold Dark Matter (CDM) type
models, have been very successful in accounting for many of the
observed properties of galaxies.  Previous investigations have focused
on the emission properties of galaxies in optical, infrared, sub-mm
and radio wavebands. Some of these successes include matching the
Tully-Fisher relationship, the optical luminosity function, the H\,I
mass function, the galaxy correlation function, and the distribution
of sizes and Hubble types, both for local and distant galaxies
\citep[most recently][]{sp:99,cole:00,dg:00,ntgy:01,spf:01,swtk:01}.

However, with the notable exception of \citet{kauf:96}, these models
have not been used to address observations of absorption systems in
much detail, though several papers have performed the basic
consistency check that the total amount of cold gas in these models is
at least as much as that observed in damped Lyman-$\alpha$ (DLA)
systems \citep{baugh:98,spf:01}.  The importance of utilizing
absorption system data to confront galaxy formation models should not
be underestimated.  The optical properties of galaxies are determined
by a sequence of poorly understood physics: star formation, the
initial mass function, supernova feedback, metal enrichment, dust,
etc. In contrast, absorption systems present an orthogonal means of
observing the high redshift Universe, and are subject to different
theoretical uncertainties as well as different observational selection
effects. Demanding that a single theory should be able to account
simultaneously for the observed properties of {\emph both} absorption
and emission systems therefore presents a stringent test for any
theory of galaxy formation.

In this paper, we use a model that has previously been shown to
successfully account for the properties of nearby and $z\sim3$
galaxies identified in emission \citep{sp:99,spf:01} to develop a
parallel picture that can also account for the observed properties of
absorption systems. This paper is the second in a series based on this
approach. In the first paper \citep[hereafter paper I]{mpsp:01}, we
demonstrated that the kinematic properties of DLA systems (absorption
systems with H\,I column densities $\geq 2 \times 10^{20} \cm2$) as
traced by low-ionization state gas \citep{pw:97,pw:98} could be
understood in this context if the gaseous disks of galaxies are very
extended, so that a significant fraction of the absorption systems
arise from lines of sight passing through multiple disks. We showed
that the metallicity and number density of DLA absorbers in this model
were also in general agreement with observations.  In this paper, we
turn our attention to the kinematics of the associated high-ionization
state gas and its correlation with the cold neutral gas. The study of
the relationship between high- and low-ionization state gas in DLA
systems has been pioneered by \citet{wp:00a,wp:00b}, who found that
none of the models they explored could successfully match their
observations.

The basic premise of our model is to associate the highly ionized gas
that gives rise to CIV absorption with the hot gas in dark matter
halos and sub-halos.  Locally, high-ionization state gas like CIV is
known to be associated with galactic halos \citep{clw:01}.
Unfortunately, modelling such systems is rather complicated because the
gas is likely to be multi-phase \citep[see for example][]{mm:96}.  We
do not attempt to model the multi-phase medium in any detail, but
rather to ascertain if the amount of hot gas in the galaxy formation
model, distributed in a reasonable manner, can produce the observed
high-ion kinematics.  We find that such a model is fairly successful,
suggesting that the general picture is compatible with the combined
constraints from the high-ion and low-ion data.

We then explore the contribution of hot and cold gas in our model to
lower column density HI systems.  We find that cold neutral gas
produces sub-DLA systems (column densities of $\sim 4 \times 10^{19} -
2 \times 10^{20} \cm2$) and we demonstrate how the kinematics of these
systems may differ from the higher column density DLAS.  We also find
in our model that CIV systems with associated low ion absorption have
markedly different kinematic properties than those which do not, i.e.,
kinematic information discriminates between DLA and sub-DLA systems.

Lastly, we find that hot halo gas can produce no more than one third
of observed Lyman limit (LL) systems, implying that the majority of
these systems are produced by either uncollapsed gas or mini-halos
\citep[e.g.][]{am:98}.  We model the contribution of mini-halos and
find that when we adopt a realistic gas distribution and include an
ionizing background, we find that these halos are unlikely to be the
source of a large fraction of LL absorbers.

The paper is organized as follows.  Section \ref{sec:back} reviews the
observations of DLA systems.  Section \ref{sec:model} describes the
basic ingredients of our model. In Section \ref{sec:res}, we compare
our model with the observations. In Section \ref{sec:sub}, we
investigate the properties and origin of LL and sub-DLA systems in our
model and identify future observational tests of the picture we have
proposed.  We close with our conclusions in Section \ref{sec:conc}.

\section{Background}
\label{sec:back}
Early on it was established that damped absorption systems are an
important probe of galaxies at high redshift \citep{wolfe:86}.  The
numerous subsequent studies of DLAS have focussed on their frequency,
metallicity, and kinematics. A recent study of the frequency of DLAS
is presented by \citet{pero:01a}, and \citet{pw:02} have presented the
most recent study of their metallicity. Most of the kinematical data
has been obtained by Wolfe and Prochaska in an ambitious program using
the HIRES spectrograph \citep{Vogt:92} on the Keck Telescope. They
acquired velocity profiles for a number of metal lines associated with
each DLA system. They found that low ionization state lines 
(e.g. SiII, AlII, FeII) 
traced one another well and therefore presumably also the cold HI gas
\citep{pw:96}.  Higher ionization state lines (e.g. SiIV, CIV) showed 
different but correlated behavior, implying that they traced a different 
phase of the gas \citep[see also][]{lu:96}.

\citet{pw:97} introduced four statistics to characterize the gas
kinematics: $\delv$, the width containing $90\%$ of the optical depth;
$f_{mm}$, the distance between the mean and the median of the optical
depth profile; $f_{edg}$, the distance between the highest peak and
the mean, and $f_{2pk}$, the distance of the second highest peak to
the mean of the profile.  The last three statistics are suitably
normalized to have ranges between $0$ and $1$ or $-1$ and $1$.  They
also introduced three statistics to quantify the correlations between
the high- and low-ion kinematics \citep{wp:00a}.  These are
\begin{itemize}
\item $\delta v$, the offset between the mean velocity of the high- and 
low-ionization state gas,
\item $\frto$, the ratio of the velocity widths of the high- and 
low-ionization state gas, and
\item $\zeta(v)$, the cross-correlation between the two gas states.
\end{itemize}

Previous semi-analytic modelling within the CDM cosmogony by
\citet{kauf:96} suggested that the majority of the cross section to
DLA systems should come from small mass halos with virial velocities
between 35 and 50 $\kms$. \citet{pw:97} investigated a number of
models to explain the kinematics of the low ionization state gas. They
concluded that the distribution of velocities predicted by
\citet{kauf:96} was strongly ruled out by the data. The only model
they investigated that was compatible with the data was one in which
the absorption systems arose from a thickened, rapidly rotating disk
($V_{\rm rot} \simeq 200 \kms$). \citet{jp:98} demonstrated that most
CDM-based models for galactic disks were incompatible with the
observed velocity widths of the DLA systems --- under the assumption
that a single disk gave rise to each absorption system. Other workers,
however, found that multiple component models, in which some of the
DLAS were produced by lines of sight passing through multiple disks or
proto-galactic clumps, could explain the low-ion observations within
the context of hierarchical structure formation \citep[paper
I;][]{hsr:98,mm:99}.

\citet{gard:01} used cosmological hydrodynamical simulations to
determine the cross section of DLA and LL systems.  They found their
simulations could produce the observed number density of DLA and LL
systems in a $\Lambda$CDM cosmology if sufficiently small mass halos
were considered ($60 \kms$ for DLA systems and $30 \kms$ for LL
systems). They found the cross section as a function of halo virial
velocity could be fit by a power law with a slope of $\approx
1.6$. However, \citet{pw:01} have pointed out that combining this
distribution of cross-sections with the velocity width to virial
velocity relationship found by \citet{hsr:98} and in paper I is not
compatible with the kinematic data.  A slope of $2.5$ is needed to
produce the correct distribution of velocity widths \citep{hsr:99}
which is consistent with what we found for our model in paper I.  The
different value found by \citeauthor{gard:01} is probably due to the
fact that there is effectively no feedback in their simulations
leading to much higher baryon fractions in small mass halos.  In
models with efficient feedback we would expect the contribution from
these low mass halos to be substantially decreased.

\citet{wp:00b} found that none of their models {\it including the
thick disk model} were consistent with the high-ionization state gas
kinematics.  They suggested that the multiple component model might be
more successful, which we now explore.

Another long-standing question pertaining to QSO absorption systems is
their connection with the galaxy population identified in emission.
At $z\la 1$, there is compelling observational evidence that metal
line and LL absorbers are associated with galaxies
\citep{steid:97,clw:01}.  A theoretical model has been proposed by
\citet{mm:96} wherein these systems are produced by gas clouds in the
halos of galaxies.  At higher redshifts, it has been proposed that many
of the absorbers may reside in mini-halos \citep{am:98}. We also
investigate the origin of lower column density systems in the context
of our model.

\section{The galaxy formation model}
\label{sec:model}
For our analysis we use the semi-analytic model described in
\citet{sp:99} and \citet{spf:01}. The backbone of the model is a Monte
Carlo realization of a ``merger tree'', which represents the build-up
of halos over time through merging and mass accretion
\citet{sk:99}. Initially, the hot gas is assumed to be distributed
like the dark matter (here, a singular isothermal sphere), to be
uniformly at the virial temperature of the halo, and not to have any
substructure. The cooling radius (the radius within which gas is dense
enough to have cooled) is then calculated using the radiative cooling
function for atomic gas in collisional equilibrium. When a halo
contains more than one galaxy, gas is assumed to cool only onto the
central object.  Recent studies \citep{yssw:02,hell:02} suggest that
this simple model is in good agreement with cosmological
hydrodynamical simulations, at least in the absence of feedback. All
gas in these simplified models is labelled as either ``hot'' (at the
halo virial temperature) or ``cold'' ($T \la 10^4$ K).

When halos merge, the central galaxy of the largest progenitor halo
becomes the new central galaxy, and all other galaxies become
satellites. Satellites lose angular momentum due to dynamical friction
and gradually fall towards the center of the halo, where they may
eventually merge with the central object. 

Cold gas is converted to stars using a simple empirical recipe, at a
rate proportional to the total mass of cold gas in the galaxy. Several
different recipes were investigated in \citet{sp:99} and
\citet{spf:01}, but here we adopt the ``collisional starburst''
recipe, which was shown to produce the best agreement with the
observed galaxy population at $z\sim3$, and with the total mass of
neutral hydrogen implied by observations of DLAS \citep{spf:01,pws:01}. 
In this model, cold gas in isolated disks is assumed to form stars with a
low efficiency (similar to that in nearby observed spiral galaxies),
while a more efficient starburst mode is triggered by galaxy
mergers. Gas that has settled into a disk may be heated and removed by
supernova (SN) feedback. The efficiency of the SN feedback is assumed
to be inversely proportional to the potential well depth of the
galaxy. Each generation of star formation produces a certain yield of
heavy elements, which are mixed with the cold gas or ejected by SN
into the hot gas halo.

After reionization, cooling only takes place in dark matter halos with
virial velocities $\vvir \ga 40 \kms$\footnote{The virial velocity is
just the circular velocity at the virial radius, which for isothermal
spheres is the circular velocity everywhere.} because of suppression
of gas collapse and cooling by the UV background.

All of our analysis has been done assuming the currently favored
$\Lambda$CDM cosmology \citep[e.g.][]{prim:02} with
$\Omega_{\Lambda}=0.7, \Omega_{M}=0.3, \Omega_b=0.038, h=0.7$, and
$\sigma_8=1.0$ and at a redshift of three. The values of the free
parameters that control the efficiency of star formation and SN
feedback, the chemical yield, etc., are as adopted in \citet{spf:01}
based on low redshift galaxy observations.

Below we describe in some detail how we incorporate a model for both
the low- and high-ionization state gas in absorption systems into this
picture. A schematic illustration of the model is shown in Figure 1.

\begin{figure} 
\centering 
\vspace{-40pt} 
\epsfig{file=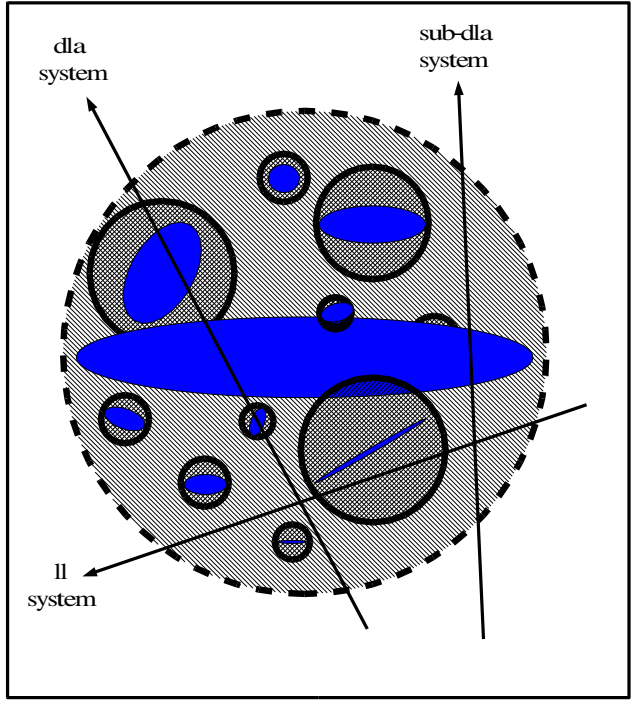,width=\linewidth} 
\vspace{-20pt}
\caption{Gas in a typical halo at $z=3$ seen in projection.  The
filled ellipses are cold neutral gas disks, while the solid circles
mark the edge of the spherically distributed hot gas in subhalos.  The
entire halo is also filled with hot gas out to the virial radius (the
dashed line). Three lines of sight through the halo illustrate how
different components can give rise to DLA, sub-DLA and LL systems in
our model (see section \ref{sec:sub}).  }\label{fig:pic}
\end{figure}

\subsection{Low-Ionization State Gas}

We assume that the low-ionization state absorption is associated with
the ``cold'' gas component of the galaxy formation model. In paper I,
we explored a number of models for the distribution of cold gas in
each galaxy and found that only by having very extensive gas disks was
it possible to produce enough overlapping cross-section for a line of
sight to a distant quasar to pass through multiple disks and thus
produce low-ion kinematics in agreement with the observations.  We
found that a model where the cold gas was distributed such that its
surface density decreases proportionally to $R^{-1}$ until a
truncation radius $R_t$ of column density $N_t = 4 \times 10^{19}
\cm2$, produced absorption systems consistent with the low ion
kinematics observed by \citet{pw:97,pw:98} as well as with the
observed $f(N)$ distribution\footnote{$f(N)$ is the number of
absorption systems with column densities between N and N+dN found per
comoving path length.}  and metallicities of DLA systems (see paper I
for more details).

We take the vertical ($z$) distribution of the cold gas to be
exponential with a scale length $z_s$ such that the disks are thin.
Thus the gas distribution for each galaxy is given by 
\beq
\label{eq:mestel} \rho(R,z)= {{N_t}\over{2z_s}} {{R_t}\over{R}}{\rm
e}^{-{{|z|}/{z_s}}}
\hspace{10pt} (R < R_t).
\eeq
We distribute the cold gas in 10 clouds to properly sample the velocity 
field.  Each cloud is given a small, 1D random velocity of 
$\sigma_{low}=10 \kms$.

We refer to the cold gas as being in ``disks'', and we implicitly
assume that the cold phase is concentrated and rotationally supported
when implementing the star formation recipes. However, for the
purposes of the kinematic modelling, the cold gas need not be
rotationally supported --- by disks we simply mean flattened systems.
We emphasize that this description is a crude simplification of the
distribution and dynamics of cold gas in real galaxies, especially in
interacting galaxies, which show rather complex behavior
\citep{hibb:00}.  In the spirit of semi-analytic models, it is our hope
that such a simplified description can still capture the general
behavior of the cold gas when averaged over a large number of systems.

\subsection{High-Ionization State Gas}

The high-ionization state absorption is assumed to arise from the hot
gas in halos. In the standard galaxy formation model, all the hot gas
is assumed to be at a single temperature and only a single gas phase
is considered.  These simplifications may be reasonable for
calculating the cooling rate of the gas, but for describing kinematics
we will need to consider the substructure of the hot gas, and we will
assume that the majority of it is in a phase suitable for CIV
absorption, in general at temperatures less than the virial
temperature of the halo.  We will continue to refer to this as hot gas
because it is dynamically hot even if parts of it have cooled.

To compare to the observations we must convert the hot gas mass to a
CIV column density, $N_{CIV}$.  The column density of CIV along a line
of sight is related to the column density of hydrogen by 
\beq
N_{CIV}=\fciv Z_{hg}N_{H}, 
\eeq
where $Z_{hg}$ is the metallicity of the gas and $\fciv$ is the fraction of 
the gas in a state that produces CIV absorption. The column density in 
HI is related to the total amount of hydrogen by $N_{HI}=(1-x)N_{H}$, 
where $x$ is the ionization fraction of the gas.  The quantities $x,\fciv$ and
$Z_{hg}$ can vary locally in the gas, but we will assume that suitably 
averaged quantities can be defined to give the global relationships we use 
here.  

We emphasize that only the metal lines in the systems are measured,
which creates an inherit degeneracy between the total amount of gas,
its metallicity and its ionization state.  For simplicity we will
treat the ionization state and metallicity to be uniform within each
halo and therefore $N_{CIV}$ is directly proportional to $N_{H}$.  The
reader should bear in mind however that an increase in $N_{H}$ can
equally well be thought of as an increase in $\fciv$ or $Z_{hg}$.  We
well return to this issue in \S~\ref{sec:conc}.

For simplicity we take $\fciv=1$, noting that a smaller value of
$\fciv$ can be directly offset by an increase in the hot gas
metallicity. We fix $Z_{hg}$ to be $-1.5$~dex for all hot gas in all
halos, which produces a good match to the CIV column density
distribution in DLA systems as measured in \citet{wp:00a}. The mean
metallicity of DLA systems at $z > 2$ is $\approx -1.5$ to $-1$ dex
\citep{pw:02}, which is consistent with our assumed $Z_{hg}$.  However,
because of the degeneracy mentioned above, this value is directly
proportional to our choice of $\fciv$. Thus this can be thought of as
a minimum metallicity since $\fciv \le 1$.  A realistic estimate of
$\fciv$ might be $\approx 0.35$ which would then require $Z_{hg}
\approx -1.0$~dex.  Alternatively, the hot gas metallicity could be
lower if there is more hot gas in halos than calculated by our simple
recipes.

The galaxy formation code calculates the metallicity of the hot gas as
part of the chemical enrichment model described briefly above.  We
find that if we use the metallicities calculated in this way (which
typically show a large scatter at fixed halo mass), we obtain too
large a spread in the CIV column density distribution. This is not too
disturbing, as the modelling of the reheating and ejection of metals
by supernova feedback is extremely uncertain. The galaxy formation
model was calibrated to produce the correct mean observed stellar,
cold gas, and hot gas metallicities in $z=0$ galaxies and clusters,
but this is no guarantee that it produces the correct results for
these quantities at $z=3$, or the correct dispersion in the
mass-metallicity relation.

However, it is also possible that correlations between $\fciv$,
$Z_{hg}$ and the gas density conspire to reduce the spread in observed
CIV column densities.  An example of this kind of conspiracy would be
a starburst which produces high metallicity in the hot gas but also
reduces the hot gas column density or changes the ionization state of
the gas such that CIV is no longer the preferred ionization state of
carbon. Unfortunately, the large number of uncertainties mean that few
constraints can be placed on any aspect of the modelling that goes
into determining CIV column densities. We discuss constraints that can
be placed on these quantities in section \ref{sec:res}.

We will assume that the hot gas is distributed like the dark matter,
i.e.\ density proportional to $r^{-2}$. For our kinematic study we
would like to consider hot gas associated with each galaxy in the
halo, while in the galaxy formation model, all hot gas is assumed to
be associated with the central galaxy.  To do this we assume that the
hot gas in each subhalo is related by a parameter $f_{sub}$ to the
total amount of dark matter in the subhalo.  Thus the mass of hot gas,
$m_{hg}$ in a given subhalo is 
\beq
m_{hg}=f_{sub}m_{dm}{{M_{hg}}\over{M_{dm}}} 
\eeq 
where $m_{dm}$ is the
dark matter mass of that subhalo and $M_{hg}$ and $M_{dm}$ are the
total mass in hot gas and dark matter respectively, summing over all
subhalos and the parent halo.  The remaining hot gas is then
associated with the parent halo.  In our realizations we find that
subhalos can contain up to $2/3$ of the halos mass, requiring
$f_{sub}$ to have values between $0$ and $1.5$.  In the case that
$f_{sub} = 0$, there is no hot gas associated with the subhalos and
our high-ion model is similar to the one explored in \citet{wp:00b}.

We note that this is far from a model of the hydrodynamics of the hot
gas in the halo, which is rather difficult to model especially
including the poorly understood physics of supernova feedback. A
successful model of gas in the halo of the Milky Way, where there is
plentiful data, still eludes researchers \citep{cbr:02}, so we must
accept that at the present moment we can only hope to sketch a
rudimentary framework of the situation at high redshift.

For the purpose of simulating spectra we assume that a fraction
$\fciv$ of the hot gas is no longer at the virial temperature but has
cooled into pressure confined clouds and is photoionized with a
temperature $\approx 5 \times 10^4 \deg$K.  We use 15 clouds to
properly sample the velocity field and assume clouds have random
velocities, $\sigma_{CIV}$, proportional to the halo's circular
velocity (see Table \ref{tab:mod}).

We compare our model only to the CIV data which best samples the
hottest phase of the gas.  The fact that there are some differences in
the high-ion kinematics between ions \citep[e.g.\, Si\,IV vs.\, C\,
IV]{pw:02} implies variations in the ionization state or metallicity
of the gas.  Future modelling of this may help us to understand
metallicity, density and ionization gradients in the hot gas.

\subsection{Mini-halo Model}
Mini-halos ($\vvir \le 35 \kms$) are usually not considered in discussions 
of galaxy formation because their expected luminosities make them difficult
to observe.  However, \citet{am:98} have shown that there may be sufficient
cold gas in mini-halos at high redshift to comprise the LL systems at those
redshifts.

Recently the issue of mini-halos has also been widely addressed
because of what has been named the {\it dwarf satellite problem}; the
fact that N-body simulations expect hundreds of mini-subhalos in a
Milky Way sized halo \citep{klypin:99b,moore:99a}, but only a couple
dozen or fewer are observed.  While many exotic solutions to this
problem have been proposed, the simplest is that the extragalactic UV
background suppresses the accretion and cooling of gas in these
mini-halos, and therefore stars form in only the small number that
managed to collapse before the epic of reionization.  Models of this
scenario \citep{bkw:00,benson:02,somer:02} have been fairly successful
in explaining why only a small fraction of mini-subhalos would have
formed stars.

To explore the contribution of mini-halos to LL systems we use the
{\it squelching} model of \citet{somer:02} to determine the amount of
hot and cold gas in mini and low-mass ($35 \kms > \vvir > 50 \kms$)
halos.  This model uses a fitting function found by \citet{gned:00} in
hydrodynamical simulations to determine the amount of gas that can
accrete onto a dark matter halo in the presence of an ionizing field.

We assume that gas will only contract to density profiles steeper than
the dissipationless dark matter.  Therefore to explore the {\bf
maximal} contribution of these halos to LL systems, we take their
projected density profiles to go as $R^{-1}$ like the cold gas disks
and hot gas in galaxies described above (equation \ref{eq:mestel}).
We truncate the gas at the virial radius or we consider cases where
the gas is truncated at a specified column density, $N_t$.  In the
latter case the truncation radius can then be evaluated by \beq R_t =
\sqrt{M_{gas} \over {2 \pi N_t m_H}} \eeq where $m_H$ is the mass of
the hydrogen atom.

\begin{figure} 
\centering 
\vspace{10pt} 
\epsfig{file=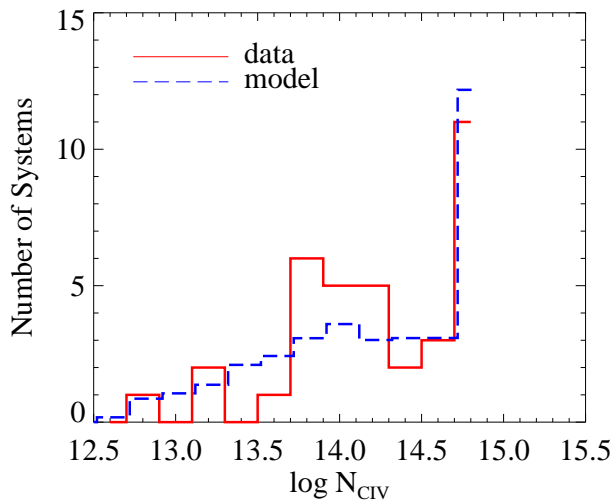,width=\linewidth} 
\vspace{5pt}
\caption{The distribution of CIV column densities for the data and the model
are shown.  The model is normalized to have the same number of systems as the
data so that Possion errors can easily be estimated.  The C\,IV doublet 
is saturated for systems with log column densities higher than $14.8$; these 
systems are binned together. We have chosen a hot gas metallicity of $-1.5$ 
dex so that the low-end cutoff of the model matches the data.
}\label{fig:civ}
\end{figure}

\section{Results}
\label{sec:res}

We pass multiple random lines of sight through each halo to produce
Monte-Carlo realizations of spectra in the manner described in paper
I.  Enough realizations are performed to give us at least 10,000 sight
lines with DLA systems in them.

Figure \ref{fig:civ} shows the distribution of CIV column densities
associated with DLA systems produced in our model. We see that the
model distribution is in good agreement with the observations. We
chose the hot gas metallicity ($-1.5$ dex) so that the bottom range of
log column densities is about $12.5$ as seen in the data.  The highest
bin contains all systems with log column densities greater than $14.8$
as this is where the C\,IV doublet becomes saturated and only lower
limits on the column density can be determined.  Of course we expect
that there is some spread in the hot gas metallicities. If we take
$Z_{hg}$ to be distributed as a Gaussian in log metallicity then a
mean of $-1.5$~dex and a standard deviation of $0.5$ is in equally
good agreement with the observations.

\begin{figure} 
\vspace{-10pt} 
\epsfig{file=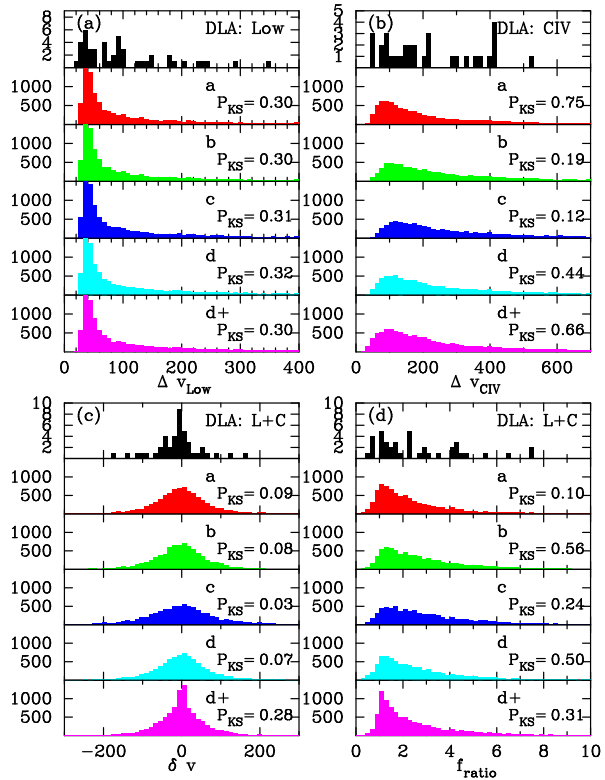,width=1.\linewidth} 
\vspace{-5pt}
\caption{The four most discriminatory tests are shown for high and low ions.
These are the $\delv_{low}$, $\delv_{CIV}$, $\delta v$ and $\frto$ tests.
Results are shown for four models where the parameters $\sigma_{CIV}$ and 
$f_{sub}$ have been varied (see Table \ref{tab:mod}). Also included 
is a model (d+) where we have added an additional source of CIV absorption
associated with the cold gas disk. All the models are consistent with the 
observations. Increasing $f_{sub}$ and the addition of some CIV absorption 
in the gas disk improves the models' agreement, but not drastically.
Adjusting other parameters like $\sigma_{CIV}$ can improve one 
statistic ($\delv_{CIV}$ in model b) but at the expense of another
statistic ($\frto$).  
}\label{fig:kin}
\end{figure}

We can place a lower limit on the amount of hot gas in galactic halos
needed to match the $N_{CIV}$ distribution by taking the highest
reasonable values for $\fciv$ and $Z_{hg}$.  Since we have already
started with the maximum value of $\fciv$, the only way to lower the
amount of hot gas is to increase the metallicity, $Z_{hg}$. If we
believe that $Z_{hg}$ must be less than a third solar then we could
get the same results with one tenth the mass of hot gas.  Any model
with less than this amount of hot gas would have difficulty producing
enough CIV absorption.

Figure~\ref{fig:kin} shows the results for the most important
kinematic characteristics of the high-ion gas: $\delv_{low}$,
$\delv_{CIV}$, $\delta v$, and $\frto$ (see Section~\ref{sec:back} for
definitions of these quantities). We show the results for five
variants of our model (a-d, and d+), summarized in Table
\ref{tab:mod}. In the first four models we vary the parameters
$\sigma_{CIV}$ and $f_{sub}$ as detailed in Table \ref{tab:mod}. All
four models produce acceptable fits to the kinematic data and none of
them can be rejected with high confidence.  The statistical tests are
sensitive to complimentary aspects of the DLA kinematics and place
relatively tight constraints on the model parameters.  For example,
increasing $\sigma_{CIV}$ (model b) improves the $\frto$ but lowers
the value of the $\delv_{CIV}$ statistic. Making the gas more clumped
in subhalos (model d) improves the agreement with the observations but
we cannot rule out a model with no clumping (model c).

Previous attempts by \citet{wp:00b} to model the high-ion gas of the
DLA failed for scenarios involving a cold, rotating disk surrounded by
a hot gas halo.  In short, the rapidly rotating disks favored by the
low-ion kinematics lead to a bimodal distribution of $\delta v$ values
which is ruled out at a high confidence level by the DLA observations.
The source of this inconsistency is the fact that the hot gas velocity
field is centered at the rest-frame velocity of the halo while the
rotating cold gas has a preferred velocity of $\sim \pm \vvir/2$.
This offset between the cold disk gas and hot halo gas is not found in
the observed $\delta v$ distribution.  For the disks in our model,
this offset also applies.  The key differences between our model and
the single disk models with respect to this issue are:

(1) the rotation speeds of the SAM disks are 
considerably smaller than the $\vvir \sim 200$~km/s required by
single disk scenarios \citep{pw:98}; and

(2) the centroids of the hot and cold gas velocity fields converge
as the number of disks intersected along a sight line increases.

\begin{center}
\begin{table} 
\begin{tabular}{ccccccc}
\hline
Model & $\sigma_{CIV}$ &  $f_{sub}$ & $\delv_{low}$ & $\delv_{CIV}$ &
$\delta_v$ & $\frto$\\
\hline
\hline
a   & $\vvir/\sqrt{2}$  & 1.0 & 0.30  & 0.75 & 0.09  & 0.10\\	  
b   & $\vvir         $  & 1.0 & 0.30  & 0.19 & 0.08  & 0.56\\	
c   & $\vvir         $  & 0.0 & 0.31  & 0.12 & 0.03  & 0.24\\
d   & $\vvir         $  & 1.5 & 0.32  & 0.44 & 0.07  & 0.50\\
d+  & $\vvir         $  & 1.5 & 0.30  & 0.66 & 0.28  & 0.31\\
\hline
\end{tabular}
\caption{Five models where the parameters $\sigma_{CIV}$ and $f_{sub}$
are varied.  In the d+ model CIV absorption associated with the cold
gas disk is included.  The last four columns give the KS probability
that the model is consistent with the data.  All five models produce
acceptable matches to the data.  The d+ model produces significantly
better agreement with the $\delta_v$ statistic than the other models.
}\label{tab:mod}
\end{table}
\end{center}

\begin{figure} 
\centering 
\vspace{5pt} 
\epsfig{file=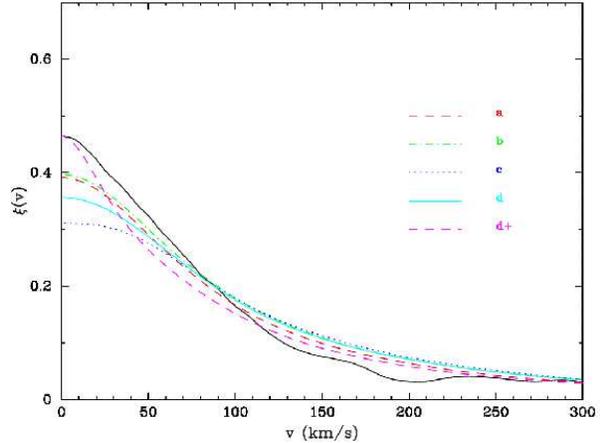,width=\linewidth} 
\vspace{5pt}
\caption{The cross-correlation, $\zeta(v)$, between high- and low-
ionization state gas.  The solid line is for the data while the other
line types are for the five models listed in Table \ref{tab:mod}.  We
see that the first four models show good agreement with the data down
to velocities of $50 \kms$ but then become less correlated than the
data.  By adding a high-ion component that rotates with the low-ion
gas (model d+), we increase the correlation at small velocity
differences, obtaining a better fit to the observed $\zeta(v)$ and
leading to an improvement in the $\delta_v$ test.  }\label{fig:cor}
\end{figure}

While a large range of parameters (Table \ref{tab:mod}) lead to
acceptable models, the agreement with the $\delta_v$ statistic is
marginal.  This can be attributed to the relatively large number of
observed DLA with nearly zero offset between the mean velocities of
the CIV and low-ions compared to the models.  Figure \ref{fig:cor}
shows the cross-correlation between high and low ions in our models
and in the data.  We see that the four models are in good agreement
with the data when the velocity difference is greater than $50 \kms$
but become less correlated than the data for smaller velocities.  This
implies that there are additional correlations between high and low
ions that we have not including in our modelling.

We therefore hypothesize that some of the CIV is not produced by hot
halo gas, but instead is associated with the cold gas disk as is seen
in the Milky Way \citep{ssl:97}.  This component would then be highly
correlated with the low-ion kinematics.  We model this component by
taking its total log column density to be $12.5$ and its kinematics to
be identical to the low ions.  We then add this to the CIV produced by
the hot halo gas.

Figure~\ref{fig:cor} shows the cross-correlation for a model with this
additional component (model d+).  One sees that the agreement with the
data for velocity differences less than $50 \kms$ is greatly improved.
These models also have more success with the $\delta_v$ statistic
(Figure~\ref{fig:kin}) than the models without the additional
component.  Thus we conclude that our model without any additions is a
good model of the kinematics for scales larger than $50 \kms$.  At
smaller scales detailed modelling of the ISM will be necessary to truly
model the contribution of high-ion gas correctly.

Therefore we conclude that the basic picture presented here for the
origin of the high- and low- ionization state gas in absorption
systems successfully reproduces all known observable properties of
these systems (aside from the possible issue of hot gas
metallicities). Next we turn to modelling lower column density
absorption systems, about which much less is known. Here, future
observations will be able to test the validity of the scenario we have
presented for DLA systems.

\section{Sub-DLA and LL systems}
\label{sec:sub}
We have investigated the properties of DLA systems and the CIV
absorption associated with them.  Our model also makes predictions
relating to gas that is below the DLA  column density limit of
$\log(N_{HI}) = 20.3$.  It has been observed locally that gas disks
are truncated at column densities of a few $\times 10^{19} \cm2$,
which has been attributed to photoinization by extra-galactic UV
radiation \citep{cs:93,malo:93}. We have found that when we truncate
our disks at $4 \times 10^{19} \cm2$, our model produces good
agreement with the kinematic data at $z\simeq 3$.  One would expect the
column density at which gaseous disks are truncated to decrease from
$z=3$ to the present as the intensity of the UV background drops.

Systems with log column densities less than $20.3$ but greater than
$19.0$ have been referred to as sub-DLA systems \citep{pero:01a}.  We
suggest that sub-DLAS should refer to cold neutral systems like DLA
and therefore should have column densities greater than a few $\times
10^{19} \cm2$.  The cutoff is probably a function of redshift as the
strength of the UV background and/or density of the gas disks varies.
Measurements of the ion abundances in these systems should be able to
tell us where the systems make a transition from cold neutral gas to
photo-ionized gas.  For now we will adopt a value of $4 \times 10^{19}
\cm2$ as the lower limit of sub-DLA systems to be consistent with the
column density at which we truncate our gas disks in our model.

At lower column densities, LL systems must arise from something other
than cold neutral gas.  Lines of sight through galactic halos that
fail to intersect cold neutral gas may give rise to LL systems (see
Figure \ref{fig:pic}).  It is also possible that LL systems are
produced by uncollapsed gas not in virialized halos, or gas in
mini-halos.

\begin{figure} 
\centering 
\vspace{5pt} 
\epsfig{file=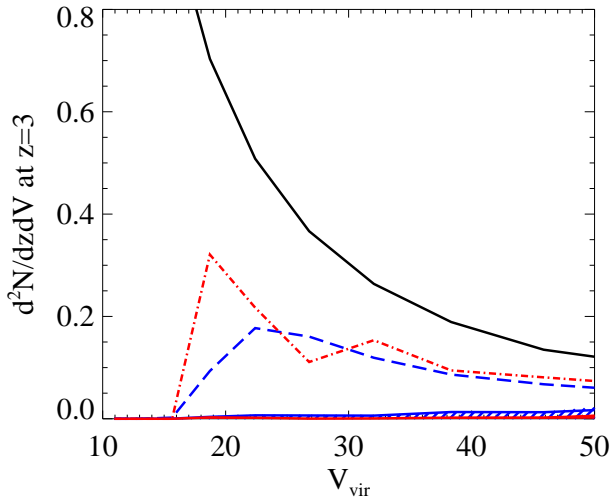,width=\linewidth} 
\vspace{5pt}
\caption{${\rm d^2N/dzdV}$ for mini-halos as a function of the halo's virial 
velocity. The solid line is the cross section of the dark matter.  
The dashed and dot-dash lines assume, respectively, that the hot or cold 
gas extend to the halo's virial radius.  The hatched area assumes the 
cold gas is truncated at $10^{19} \cm2$.  The smaller filled area assumes 
the hot gas is $99\%$ ionized.  We see that in realistic
models of mini-halos they do not contribute much cross section to LL 
absorption.  The cold gas in low mass halos may account for a 
substantial fraction of sub-DLA systems if it is truncated at a column density
lower than that of cold gas in galactic halos.
}\label{fig:mini}
\end{figure}

\subsection{The contribution from mini-halos}
Figure \ref{fig:mini} shows the number per unit redshift $\dNdz$ of
mini-halos as a function of $\vvir$. We see that the cross section for
encountering a halo with a virial velocity in the range $10-50 \kms$
at $z=3$ is 10 times as large as the observed $\dNdz$ for LL systems.

As we have discussed, halos in this velocity range will have
difficulty in accreting and cooling gas in the presence of a UV
background. However, mini-halos which collapse before reionization may
contain cold gas.  If we assume that the cold gas in these systems
extends out to the virial radius of the halo we obtain $\dNdz=3.7$
(the dashed line in Figure \ref{fig:mini}).  This is consistent with
the results of \citet{am:98}. It requires that the cold gas remain
neutral at column densities far below $10^{19} \cm2$, which is
probably unrealistic. If instead the cold gas is truncated at $10^{19}
\cm2$ then we obtain $\dNdz=0.3$, a negligible contribution to LL
systems.  In this case, the mini-halos would account for more than
half of the systems with $N_{HI} > 10^{19} \cm2$, but it is unclear
why the truncation value for mini-halos should be less than that of
galactic halos.

An examination of the hot gas in mini-halos leads to similar
results. If the hot gas is taken to be neutral and to extend out to
the virial radius then it has a rather high rate of incidence
$\dNdz=4.7$.  However, once again it is difficult to understand why
this gas would not be highly ionized.  In hydrodynamic simulations of
mini-halos in an ionizing background, \citet{ktas:99} found that at
maximum the HI column density only exceeds the Lyman limit threshold
out to a radius less than one ninth the virial radius.  If we assume
that the hot gas is uniformly $99\%$ ionized then $\dNdz < 0.06$.

Thus it seems that cold gas in mini-halos may contribute to sub-DLA
systems but that in realistic scenarios mini-halos produce relatively
few LL systems. Note that \citet{gard:01} suggested that their
hydrodynamical simulations could account for the number of LL
absorbers if they could be continued down to halos with $\vvir \sim 30
\kms$.  The reason these results are different from ours using
semi-analytic models is that the simulations have effectively no
feedback; there is much more cooling and early star formation than in
the semi-analytic models.
Also they see no effect of the UV photoionizing background on halos
with $\vvir > 50 \kms$ but cannot resolve low mass and mini-halos.
 
\begin{figure} 
\centering 
\vspace{5pt} 
\epsfig{file=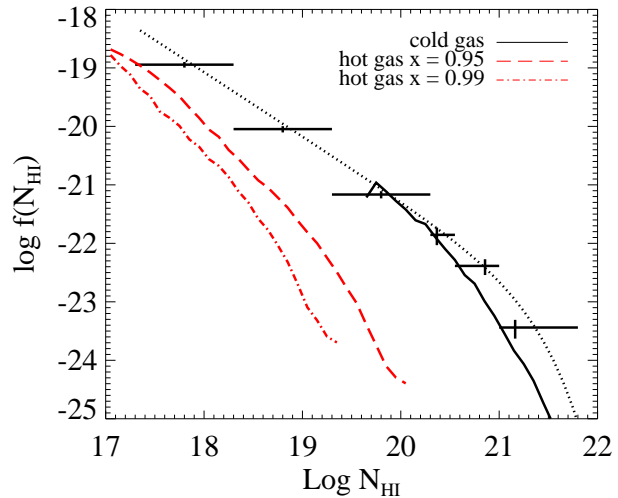,width=\linewidth} 
\vspace{5pt}
\caption{The column density distribution $f(N)$ at $z=3$ for DLA and
LL systems in our model. Also shown are the data of
\citet{pero:01a}. Note that the three points on the left are not actual
data points but an extrapolation of the dotted line
\citep[see][]{pero:01b}.  The dark line shows the column density
distribution of cold neutral gas. The dashed and dash-dot lines show
the column density distribution of hot gas assuming $95\%$ and $99\%$
ionization respectively.  One sees that there is a large drop in the
number of absorbers with log column densities between $18.0$ and
$19.8$ in our model. One also sees that hot halo gas only generates
about a third of the LL systems at $z=3$.  }\label{fig:fn}
\end{figure}

\subsection{The contribution from hot gas in galactic halos and 
uncollapsed gas}

The hot gas in halos which produces the high-ion gas in DLA and sub-DLA 
systems will also generate LL systems in lines of sight that do not 
intersect cold gas. The frequency of lines of sight that intersect hot 
gas in our model is only $0.78$ per unit redshift, well below the observed 
number density of $2.0\pm0.5$ for LL systems.  Furthermore the hot gas 
is highly ionized so that depending on the ionization fraction $x$ these 
systems may not have a high enough column density to be LL systems.  

\citet{proc:99} has observed that $x=0.97\pm0.02$ in a LL system.  
If we assume that $x$ is within this range for the hot gas in halos
then the resulting column density distribution is shown in Figure
\ref{fig:fn}.  The dashed line and the dot-dashed line show the
distribution taking $x$ of $0.95$ and $0.99$ respectively.  Column
densities have only been measured for DLA systems; \citet{pero:01b}
has extrapolated the distribution to lower column densities assuming
the distribution can be fit by a gamma function and constrained by the
total number of LL systems.  This is shown in Figure \ref{fig:fn} by
the dotted line. There is a tremendous drop (a factor of 100) in the
number of absorbers after the cold gas truncation level. Of course a
different source of absorbers may compensate for this drop, yet it
seems unlikely that they would exactly cancel the drop caused by
the truncation of cold gas disks.  More data in this regime would be
very helpful in clarifying the physical nature of the absorption
systems.

Since there does not seem to be enough cross section to produce the
observed number of LL systems in either galactic halos or mini-halos, we
must infer that the majority of LL systems are likely to be associated
with uncollapsed gas 
in the vicinity of galactic halos. This is consistent with the results
of \citet{dhkw:99}, who found that in a hydrodynamic CDM simulation,
systems with column densities of $\sim 10^{17} \cm2$, typical of LL
systems, are produced in regions with overdensities of $\sim 100$ at
$z=3$. One can see from these simulations that these systems tend to
reside in the weakly non-linear filaments surrounding halos. This gas
may be enriched by material ejected by supernovae from the halos and
will probably have velocity widths $\sim 100 \kms$.  A line of sight
passing through 1 kpc of $99\%$ ionized gas with an average
over-density of 100 at $z=3$ acquires enough optical depth to be above
the Lyman limit. These are conditions that would not be unusual around
galaxy mass halos. Further hydrodynamic simulations will be needed to
study this non-linear regime in more detail.

More data are required to identify the origin of LL absorbers at high
redshift and it is quite possible that hot gas in galactic halos, cold
gas in mini-halos, and uncollapsed gas around halos all play a
non-negligible role.  Studies of the ionization state of the gas, its
column density distribution and its kinematics will help identify the
nature of these absorption systems.

\begin{figure} 
\centering 
\vspace{10pt} 
\epsfig{file=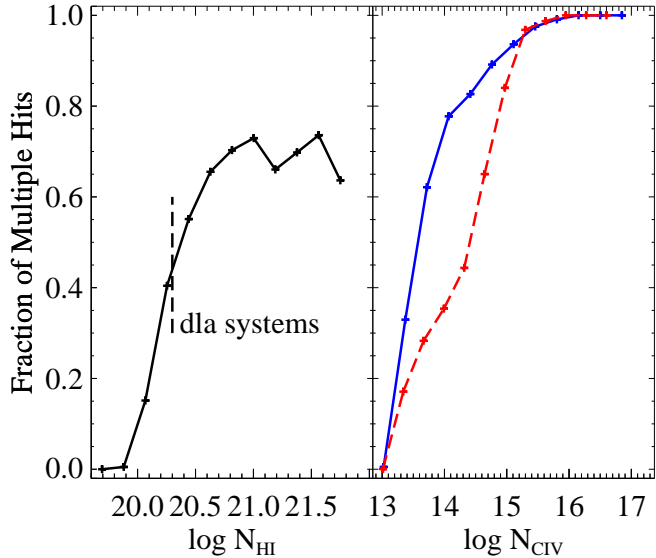,width=\linewidth} 
\vspace{10pt}
\caption{The fraction of systems composed of multiple gas disks as a
function of absorption system column density. The left panel shows DLA
and sub-DLA systems, while the right panel shown CIV absorbers.  The
DLA systems are mostly multiple component systems as required to
explain their kinematics; however the sub-DLAS (particularly with
column densities below $10^{20} \cm2$) are mostly produced by single
disks. CIV absorbers that are not associated with DLA or sub-DLA
systems (dashed line) are more likely to be single component systems
than CIV associated with cold gas (solid line).  }\label{fig:num}
\end{figure}

\subsection{The kinematics of sub-DLA systems}
\begin{figure} 
\centering 
\vspace{10pt} 
\epsfig{file=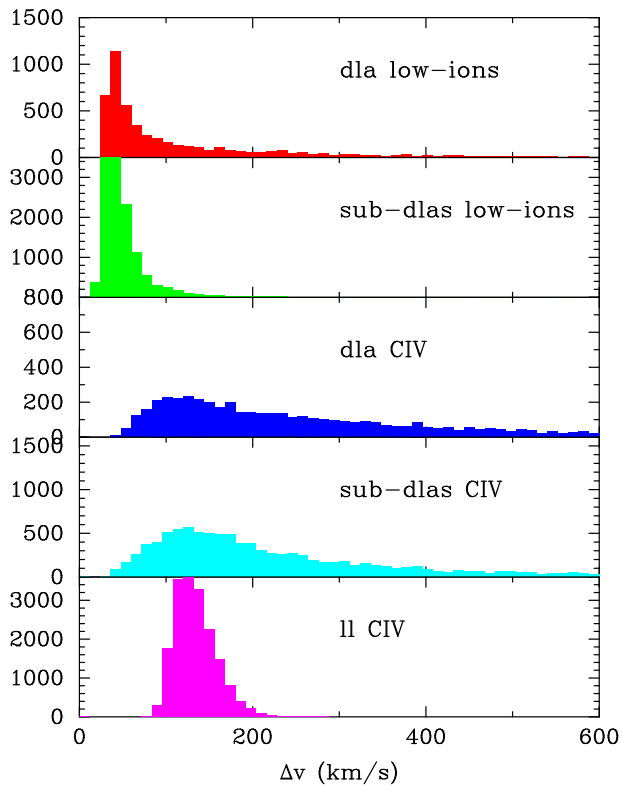,width=\linewidth} 
\vspace{5pt}
\caption{The distribution of model velocity widths $\delv$ is shown
for the low ions of DLA systems, the low ions of sub-DLA systems and
the high ions of DLA, sub-DLA and LL systems.  While the high ions
show little difference between DLA and sub-DLA systems, the low ions
are markedly different, providing a direct test of the multiple
component model.  A similar effect is seen between the high ions in
the LL systems and the others, an additional test of our model.
}\label{fig:sub}
\end{figure}

A kinematic investigation of sub-DLA systems provides a generic,
direct test of the multiple component model.  In any multiple
component model, most lines of sight pass though only a single
component. The only way for DLA systems to be dominated by multiple
components is for the column density of a single component to be below
the DLA cutoff.  This means that at some lower column density
(sub-DLA) the absorption systems must become dominated by single
component encounters.  In our model this happens at H\,I column
densities less than $10^{20} \cm2$, as shown in Figure \ref{fig:num}.
This figure shows the fraction of absorption systems produced by more
than one gas disk as a function of column density.  The sub-DLA
systems are very different below a log column density of $20$.  They
are almost entirely composed of single disk systems.  Thus we expect a
significant change in the measured kinematics for these lower column
density systems.  The difference in the measured $\delv$ for the low
ions is shown in the upper two panels of Figure \ref{fig:sub}.  One
sees the transition to many more small $\delv$ systems.  Observing the
low-ion $\delv$ in these systems should be a direct test of the
multiple component model.

\citet{omea:01} have studied a system at $z=2.5$ with a log column
density of $19.4$.  This system is neutral and thus by our definition
should be refered to as a sub-DLA system even though its column
density is less than the truncation value we use in our model.  The
system has very simple kinematics clearly indicating it is a single
component system; however, it was also selected for its simple
kinematics to study deuterium so it is unclear if it is representative
of the population.  An unbiased investigation of systems like this one
is needed to test the multiple-component scenario.

The strong trend with column density seen in our model may be an
artifact of the fact that we have rather artificially truncated all
gas disks at the same value.  A spread in values would tend to weaken
the effect; nevertheless, the trend should exist for any multiple
component model for DLA systems because such a model must always have
a larger cross-section to single encounters than multiple encounters.
Thus we believe the kinematics of sub-DLA systems is a critical test
of the multiple component model.

A similar effect would be predicted for CIV profiles if they are
composed of multiple components.  Again, lines of sight through single
components are preferentially lower column density systems.  The case
of $f_{sub}=1$ is shown the right side of Figure \ref{fig:num} for CIV
associated with DLA and sub DLA systems and for CIV with no associated
low ions. The large difference in systems with column densities
between $13.5$ and $14.5$ suggests that it may be possible to identify
which systems contain cold neutral gas from the CIV kinematics. This
provides yet another diagnostic of our model for the high ions.

The difference between the velocity widths $\delv_{CIV}$ for systems
with and without low ions is quite striking (Figure \ref{fig:sub}
lower three panels).  This trend holds even if $f_{sub} = 0$ where all
systems are produced by a single component.  In that case the DLA and
sub-DLA systems arise from lines of sight passing through the central
parts of the halo and thus are more likely to have large
$\delv_{CIV}$.  Lines of sight that do not intersect cold gas are
farther from the halo center and sample a smaller range of velocities.
We note however that since some other state of gas is giving rise to
most of the LL systems this gas may also produce CIV and thus
complicate this picture.  Further conclusions will not be possible
without also modelling the other sources of LL absorption (i.e., the
ones unaccounted for by our model).

\section{Conclusions}
\label{sec:conc}

We have presented a model for the high-ionization state gas seen in
DLA systems, based on a galaxy formation model set within the
hierarchical structure formation paradigm.  The basis of this model is
to associate hot halo or sub-halo gas with the highly ionized gas that
gives rise to CIV absorption.  Assuming a simple model for the hot gas
distribution, we generate absorption systems with kinematic properties
in reasonable agreement with observations of DLAS at $z=3$.  We thus
conclude that a CDM-based galaxy formation model can also account for
both the low- and high-ionization state gas in observed DLA absorption
systems. Agreement with the data is not a strong function of any free
parameter of the model and thus we conclude that the general concept
of a spherical hotter component as the source of CIV absorption
provides a viable explanation for the observed kinematic properties.
Associating some hot gas with subhalos and including a CIV component
in co-rotation with the low ionization state gas improves the
agreement, but models without these features cannot be categorically
ruled out.

While the galaxy formation model provides the information about the
\emph{amount} of hot and cold gas present in a given halo, our model
requires a number of additional assumptions about the gas
\emph{distribution}.  The most striking is our requirement that the
cold gas be in a rather extended configuration (paper I).  Also, in
order for our model to work, we require that the gas termed ``hot'' in
the semi-analytic model is in the form of clouds in a two-phase medium
\citep[as in ][]{mm:96}.  This is clearly beyond the level of detail
that semi-analytic galaxy formation models or hydrodynamic simulations
currently attempt to address.

Our model produces good agreement with the observed distribution of
CIV column densities under the assumption that hot gas is distributed
like the dark matter and that the fraction in a state suitable for CIV
absorption $\fciv$, and the metallicity, $Z_{hg}$, are uniform
throughout the halo.  While this description has the advantage of
simplicity, it is unlikely to be correct.  The fact that the different
high-ionization species do not always trace one another implies that
$\fciv$ and $Z_{hg}$ vary throughout the gas. 
We found that a model in which a larger fraction of the gas was
associated with sub-halos produced better agreement with the
observations, but an alternative scenario in which the subhalo gas
possesses higher values of $Z_{hg}$ or $\fciv$ might provide a better
explanation.
The other high-ionization state metal lines may be used in future
studies to help constrain the distributions of gas density,
metallicity and ionization state in high redshift halos.
	
We also explored the predictions of our model for the lower column
density sub-DLA and LL systems.  We suggest that the kinematics of
sub-DLA systems may be a decisive test of the multiple component
model.  Any multiple component model must have a large cross-section
to single component intersections.  Most sub-DLA systems should
be produced by lines of sight passing through a single disk and
therefore they should have significantly different kinematics than the
DLA systems.  This will also be true for CIV systems if the hot gas is
clumped into subhalos.  Finally, we expect a large drop in the number
of absorption systems with column densities below the level where our
gas disks are truncated, an additional test of the model.

We find that in our model, only about a third of the observed
population of LL systems are produced by hot gas in halos. 
We furthermore find in agreement with \citet{am:98} that there is enough 
cold gas in mini-halos to produce all of the LL systems at $z=3$ and that 
there is also a similar amount of hot gas in these halos.  However if 
we adopt a realistic model of the gas in
mini-halos, we find that they produce an almost negligible
contribution to the cross-section of LL systems. This is because to cover
a large area the gas must have a low column density and therefore will
be photoionzied by the UV background. Having high enough column densities
to be self shielding implies the cross-section is small and therefore
not a significant source of LL systems.

We therefore propose
that gas surrounding halos but outside the virial radius (e.g. in
filaments) may give rise to LL systems, a view supported by the
hydrodynamic simulations of \citet{dhkw:99}. If this gas is pre-enriched
or enriched by metals ejected from the halo it may also produce metal
line systems.  Further investigations using hydrodynamical simulations
will be useful in studying uncollapsed gas outside of virialized halos
and determining its contribution to absorption systems.

We have illustrated that absorption systems comprise a powerful probe
of galaxy formation. Acquiring more data of this kind, and developing
more detailed models of these systems, will help in forming a complete
picture of the properties of the gas present in the early epochs of
galaxy formation, which form the building blocks of the galaxies
that we see today.

\section*{Acknowledgements}
We thank Celine P{\' e}roux for stimulating conversations.  AHM acknowledges 
support from NASA LTSA grant NAG5-3525 and NSF grant AST-9802568.
JXP was partially supported by NASA through a Hubble Fellowship
grant HF-01142.01-A awarded by STSCI and JRP was supported by NASA 
and NSF grants at UCSC.
\bibliographystyle{mn2e}         
 
\bibliography{me,lylens,gf,abs,cosmo,reion,dm} 
\end{document}